\documentstyle[psfig]{mn}
\newif\ifAMStwofonts

\newcommand{\msun}{M$_{\odot}$}
\newcommand{\lstar}{L$^{\star}$}

\title[ERGs: ellipticals vs. starbursts]
{Extremely Red Galaxies: age and dust degeneracy solved ?}

\author[Pozzetti \& Mannucci]
	{L. Pozzetti, $^{1,2}$
	 F.~Mannucci,$^3$ \\
	$^1$Osservatorio Astrofisico di Arcetri, Largo E.Fermi 5, 50125 Firenze,
	    Italy\\
        $^2$Osservatorio Astronomico di Bologna, Via Ranzani 1, 40127 Bologna, 
            Italy (lucia@bo.astro.it) \\
        $^3$C.A.I.S.M.I.--C.N.R., Largo E.Fermi 5, 50125 Firenze,  Italy}
\date{Submitted 2 Feb 2000, Revised 28 Jun 2000, Accepted 10 Jul 2000}
\pagerange{\pageref{firstpage}--\pageref{lastpage}}
\pubyear{2000}

\begin{document}

\label{firstpage}

\maketitle

\begin{abstract}
The extremely red galaxies (ERGs) are defined in terms of their very red
optical-to-near IR colours (as R-K$>$5 or I-K$>$4). 
Originally this selection was aimed at
selecting old ($>$ 1 Gyr) 
passively evolving elliptical galaxies at intermediate redshift 
(1$<$z$<$2), but it was
soon discovered that young star-forming dusty galaxies can show similar
colours and therefore be selected in the same surveys. It is crucial to
distinguish between these two populations because they have very different
consequences on the models of galaxy formation.
Here we show that old ellipticals and dusty starburst are expected to
show different colours in the (I-K) vs. (J-K) diagram for redshift range 
1$<$z$<$2, providing thus
a useful tool to classify ERGs in large samples up to K$<$20.
This is mainly due to the fact that old galaxies at these redshifts
have a strong 4000\AA\ break at $\lambda<1.2 \mu$m ($J$ band), 
while dusty galaxies show a smoother spectral
energy distributions and therefore redder J-K colours.
We discuss this difference in detail both in the
framework of the stellar population synthesis models and by using
observed spectra. The selection criterion is also compared with the
properties of ERGs of known nature. 
We also show that this colour selection criterion is
also useful to separate the ERGs from 
brown dwarf stars showing similar optical-to-IR colours.

\end{abstract}


\section{Introduction}

The extremely red galaxies (ERGs) discovered in deep 
near-infrared (IR) and optical surveys (e.g., McCarthy et al. 1992, 
Hu \& Ridgway 1994) 
are among the most intriguing objects in the modern cosmology and
their nature is still controversial. They
are defined in terms of their very red optical/near-IR colours ($R-K>5$ or 
$I-K>4$), are very rare
at bright K magnitudes while their density approaches 0.5 $\pm$ 0.1 arcmin$^{-2}$ at K=20 (McCracken et al. 2000).
Such very red colours can be explained by two mainly opposing scenarios:
1) ERGs can be starburst galaxies hidden by large amounts of dust, or
2) can be high redshift ($z>1$) old ellipticals with intrinsically red 
spectral energy distributions (SEDs) and large positive k-corrections.
To know which population is dominant between the ERGs is a crucial test
for the models of galaxy formation: the presence of a wide-spread
population of high redshift old ellipticals
would point towards a very early formation of massive galaxies 
in a ``monolithic" scenario (Eggen, Lynden-Bell \& 
Sandage 1962, Larson 1975), 
while the presence of numerous dusty starbursts at intermediate
redshift is better fitted by the
hierarchical scenarios of galaxy formation (White \& Frenk 1991, Kauffmann, Charlot \& White 1996). 
Both these populations would have a deep impact
on the actual knowledge of the cosmic history of star formation.

The two populations can in principle be distinguished by several
observations: 
1) high signal-to-noise spectra can show the presence of emission lines,
revealing ongoing star formation activity, or, on the contrary, sharp
spectral breaks due to old stellar populations;
2) sub-mm observations can detect the far-IR emission from the hot dust
heated by the starburst; 
3) high resolution optical and near-IR images can reveal the morphology,
which is expected to be regular and centrally peaked for old ellipticals
and irregular, disturbed  or even double for dusty starbursts.
In practice these observations help only in a limited number of cases:
1) the faintness of these objects in the near-IR and especially in the
optical (typical magnitudes are K=19 and R=25) makes the
spectroscopic observations very difficult, and many Keck and VLT nights
have produced redshifts for just few objects (see below);
2) the best current sub-mm instrument, SCUBA (Holland et al. 1999), 
can still detect the far-IR flux only from exceptionally bright objects 
(e.g., Cimatti et al. 1999), while
the vast majority of the population remains below the detection threshold;
3) HST (WFPC2/NICMOS) images have been obtained only for a limited number
of objects, and also in these cases a detail optical morphological study is
hampered by the faintness of these galaxies. 
Therefore it would be useful to discover a classification tool
applicable to large samples of ERGs.

\begin{figure}
\centerline{\psfig{figure=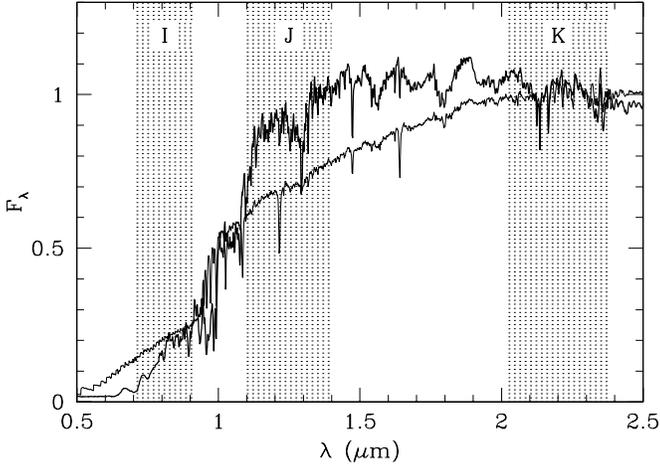,width=9cm}}
\caption{
Representative spectra from an old elliptical (thick line) and a dusty
starburst (thin line) at z=1.5. The position of the I, J and K filters
are shown. Both spectra have been computed using the Bruzual \& Charlot
(1999) GISSEL model. The thick line is a simple stellar population (SSP)
with an age of 15 Gyr, while the thin line shows the spectrum resulting from
a constant star formation rate over 1 Gyr when reddened to
E(B-V)=0.8 by a screen of dust described by the SMC law.
It is apparent as, while the I-K colour selects both class of objects,
the J-K colour can distinguish between them.
}
\end{figure}

A few ERGs were studied in great detail and can be used as benchmarks
for a classification. As expected, 
both dusty and old galaxies contribute to the ERGs population:
HR10 (Hu \& Ridgway 1994), a
colour--selected galaxy at z=1.44 (Graham \& Dey 1996),
shows a disturbed and elongated WFPC2 morphology (Dey et al. 1999) and was
also detected
by SCUBA in the sub-mm (Cimatti et al. 1998, Dey et al. 1999), 
making its classification as a dusty starburst very robust;
53W091 (Spinrad et al. 1997), a faint radio galaxy at z=1.55, whose
spectrum shows a break due to a dominant stellar population of a few Gyr old
(Heap et al.  1998); 
6 galaxies in the core of the X-ray selected cluster RXJ0848.9+4452 
(Rosati et al. 1999) at redshift 1.26 show absorption features typical
of evolved ellipticals and 3 ERGs with $K<20$  
(Cohen et al. 1998) are found to be absorption line galaxies at redshift 
$1<z<1.23$. Some ERGs still without 
redshift show no evidence for strong emission
lines (Cimatti et al. 1999, Cohen et al. 1999),  
two of them have however SEDs that require strong dust reddening.
Other ERGs can be
classified by their HST optical morphology
(Pozzetti et al. in preparation) and surface brightness profiles (Moriondo et 
al. 2000).
Finally some faint sub-mm sources have been identified with ERGs
(Smail et al. 1999).

\begin{figure}
\centerline{\psfig{figure=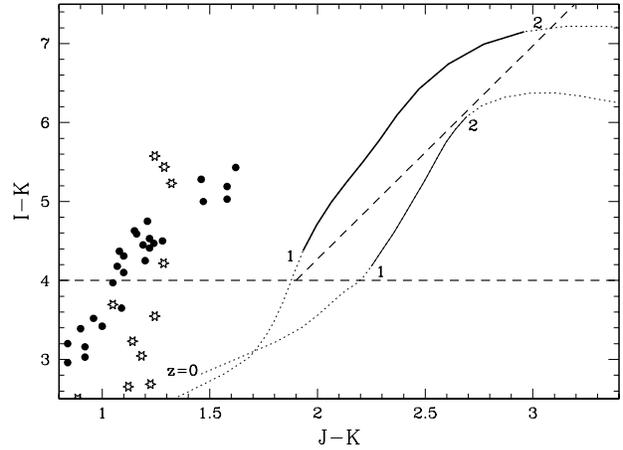,width=9cm}}
\caption{
Colors predicted for ellipticals (thick line) and starburst
(thin line). The dotted lines represent the
colours for z$<$1 and z$>$2, while the solid lines show the redshift range
1$<$z$<$2. The dashed lines indicate the ERGs selection criterion
(I-K$>$4) and separate the region where the two
populations of ERGs are expected to fall. Stars and dots show the stars
in the samples by Pickles (1998) and the brown dwarfs by Legget et al.
(1998), respectively. The I-K colours of these stars
are similar to those of the ERGs, but the two populations
show different J-K average colours, the brown dwarfs being bluer.
}
\end{figure}

All the ERGs with spectroscopic redshifts fall in the range $1<z<2$.
Moreover, most of the other ERGs discovered by the current surveys
are expected to fall in this redshift range
because of their optical and near-IR colours and magnitudes. Indeed
\lstar\ objects at higher redshifts are expected to fall below 
the magnitude threshold of the current surveys ($K<20$), 
while objects at lower redshift tent to drop out of the colour 
selection for ERGs ($R-K>5$ or $I-K>4$) because of smaller
k-corrections.
For these reasons we have studied a photometric way to distinguish
between the two populations in the redshift range 1$<$z$<$2. 
This recipe is based on optical/near-IR broad-band data only
to be easily applied to large samples of ERGs; it is derived by using 
numerous stellar synthesis models and observed spectra.
The prediction can be compared to the properties of
observed ERGs for which we can assign a 
classification based on morphological, spectral or sub-mm information.

\section{The I-K vs. J-K colour plane}

While old stellar populations are characterized by sharp breaks, especially
around 4000 \AA, young dusty galaxies show shallower SEDs in the optical range
because dust extinction does not produce sharp breaks. This is shown in
Fig. 1 where  the representative spectra of the two classes of
objects are shown at $z=1.5$. 
This feature allows us to clearly separate the two classes of objects
using a I-K vs. J-K diagram, where the former colour measures how red an
object is, while the latter measures the steepness of its SED.
Daddi (private communication) propose a similar classification scheme
based on the J-K colour only.

In Fig. 2 we show the I-K vs. J-K plane when local galaxies are
used as templates of high redshift objects:
for the ellipticals we have used the
average of 5 local ellipticals by Mannucci et al. (in preparation), while 
we have used M82 (see Silva et al. 1998) as a typical dusty starburst.  
When redshifted between $z=1$ and $z=2$, these observed spectra fall
in different regions of the plane and for a given I-K they are separated 
by about 0.3 in J-K. In this diagram, as in the following ones, we use a
Cousins' I filter and Bessel \& Brett (1988) J- and K- filters.
In this figure we have also plotted 
some cold stars and brown dwarfs
which are known to contaminate the ERGs samples selected by the
optical-to-IR colour criterion, especially at faint flux levels
when the morphological classification becomes weaker.
When the J-K colours is considered, the two populations seem to
clearly separate (see also Fig. 4), 
at least when the cold stars in the Pickles (1998)
library and the brown dwarf in the Legget et al. (1998) are considered.
Similar results are also obtained by using models and observations by
D'Antona et al. (2000) and Burrows et al. (1997). The use of the J-K
colour can therefore remove most of the foreground red stars from the
ERGs samples.

\begin{figure*}
\centerline{\psfig{figure=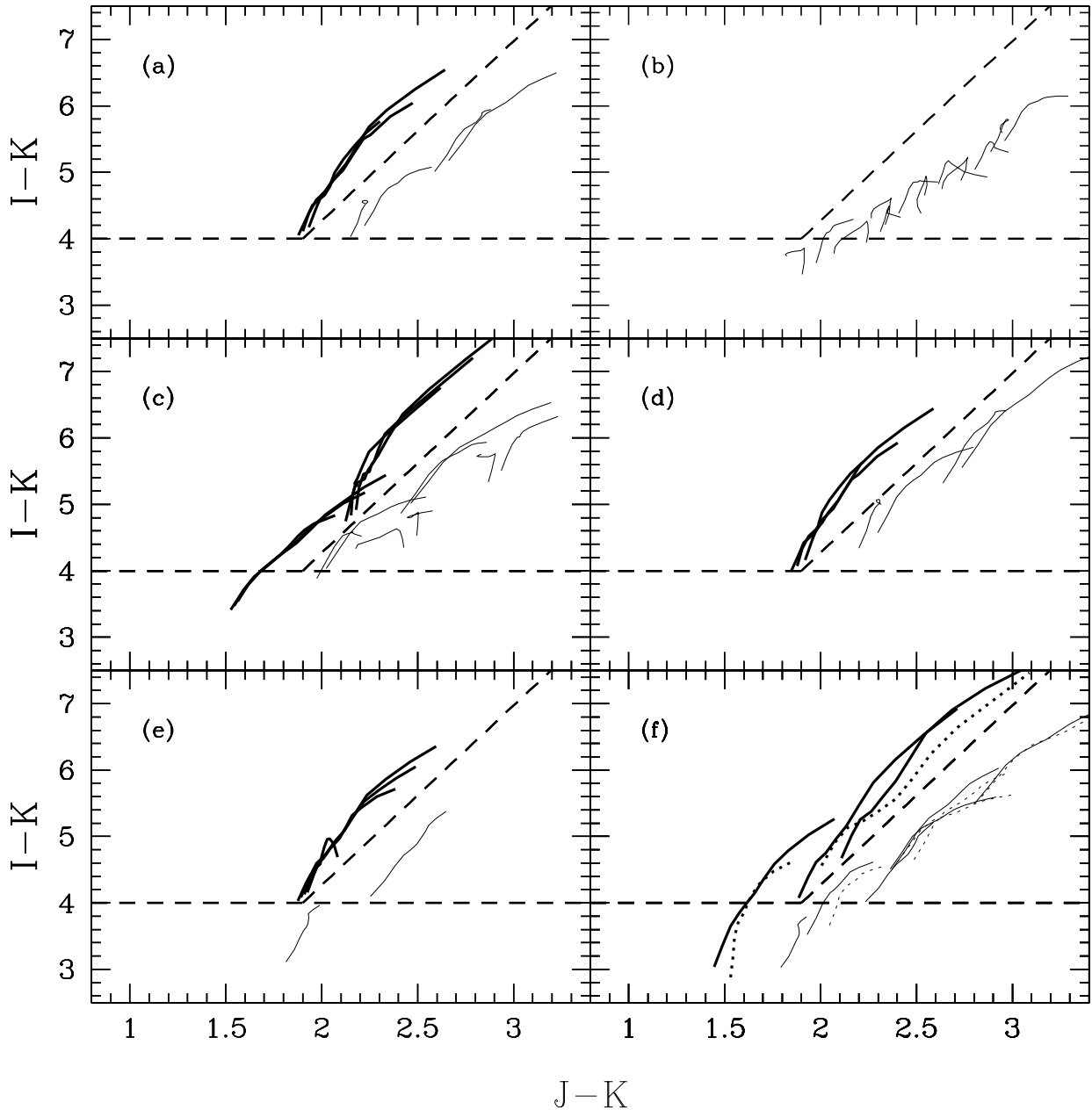,height=18cm}}
\caption{
Prediction of spectral synthesis models, described in detail in the
text. The dashed line is the same in all the panels and corresponds to
eq. (1). In all the cases, models to the left of the dashed line correspond to
ellipticals (thick lines), those to the right to dusty starburst
(thin lines).
}
\end{figure*}

In Fig. 3 we show that this separation is predicted by a large range of
models of old elliptical and dusty starburst. We have explored this
range by studying  how the results depend on
the particular spectral synthesis model used, 
the metallicity of the stellar population,
the assumed history of star formation,
the Initial Mass Function (IMF),
the assumed redshift of formation and
the properties and the distribution of dust.

The various panels of Fig. 3 show the results for 1$<$z$<$2. 
In all the cases the dashed line from Fig.  1 (see eq. 1)
separates the two populations, ellipticals on the left and starbust on
the right. We used a cosmology with $H_0=50$ km s$^{-1}$ Mpc$^{-1}$
and $\Omega=0.1$, but 
this choice is not critical for the results.
\begin{itemize}
\item Panel (a): The basic model.\\
The ellipticals are represented by simple stellar population
(SSP) of the Bruzual \& Charlot (1999) evolutionary model 
with solar metallicity (Z=0.02), 
Salpeter IMF between 0.1 and 125 \msun ~and three values of the redshift of 
formation, $z_f=3,4,6$. The starbursts are reproduced with constant
star formation rate, $z_f$=2.5 and 6, and an extinction by dust with 
$E(B-V)$=0.5 and 0.8, 
assuming the Small Magellanic Cloud (SMC) extinction law (Pr\'evot et al. 1984),
and a uniform screen external to the emitting 
region. These models reproduce the local ellipticals and starburst very well.
Despite the spectra used for Fig. 2, these spectra take into account the 
evolution of the galaxies with look-back time.

\item Panel (b): dust properties and distribution.\\
We explored the effect of 
different dust extinction laws and distributions in starbursts, in
addition to the SMC law used in the other panels. 
Firstly we have used the Calzetti (1997) law derived using starburst
galaxies, assuming, as for the SMC one,
that dust is external of the emitting region.
This may be not appropriate to reproduce starburst galaxies where dust
in supposed to be very inhomogeneously distributed and mixed with the 
emitting gas. Therefore we also used the extinction laws by 
Silva et al. (1998), Ferrara et al. (1999) and 
Cimatti et al. (1997), based on a number of different
assumptions for dust spatial distribution and optical depth,
including both absorption and scattering by dust.

\item Panel (c): metallicity effects\\
The same models of panel (a) are plotted but with two different
metallicities, Z=0.004 ($0.2$ solar, bluer curves) and Z=0.05 (2.5
solar, redder curves). The predicted colours become bluer and redder 
respectively, but
each class remains inside its region of the diagram.

\item Panel (d): IMF effects\\
Here a Scalo IMF is used, which is steeper at high masses and shallower
at low masses than the Salpeter IMF of panel (a). 
The resulting effect is a prediction of colours redder than in the basic model.

\item Panel (e): Star formation histories \\
We also changed the history of star formation by assuming
a small residual star formation for elliptical galaxies described by 
an exponentially declining SFR with a short 
e-fold time of 0.3 Gyr, and for dusty starburst an e-fold time of 0.1 Gyrs, 
at an age of 0.1 Gyr and for $E(B-V)=0.5, 0.8$.

\item Panel (f): other models \\
Independently from cosmology we plot no-evolution models for
elliptical (SSP at age=1, 5 and 15 Gyrs) and dusty starburst (constant
SFR and age=0.1,1 and 15 Gyrs and E(B-V)=0.5,0.8).
We also want to explore how the prediction change when 
different spectrophotometric models are used.
We use Barbaro \& Poggianti (1997) models (dotted lines), 
SSP for elliptical galaxies at age of 1, 15 Gyrs
and constant star formation models 
at age=1, 15 Gyrs with an extinction of E(B-V)=0.5,0.8 for starbursts.
\end{itemize}

In conclusion, by changing the various parameters of the stellar
synthesis models it is poSsible to obtain quite different spectra, but
the separation between the two populations remains a valid prediction.
Since some models fall near the separation line,
some degree of contamination between the two class could be present
near it.

\begin{figure}
\centerline{\psfig{figure=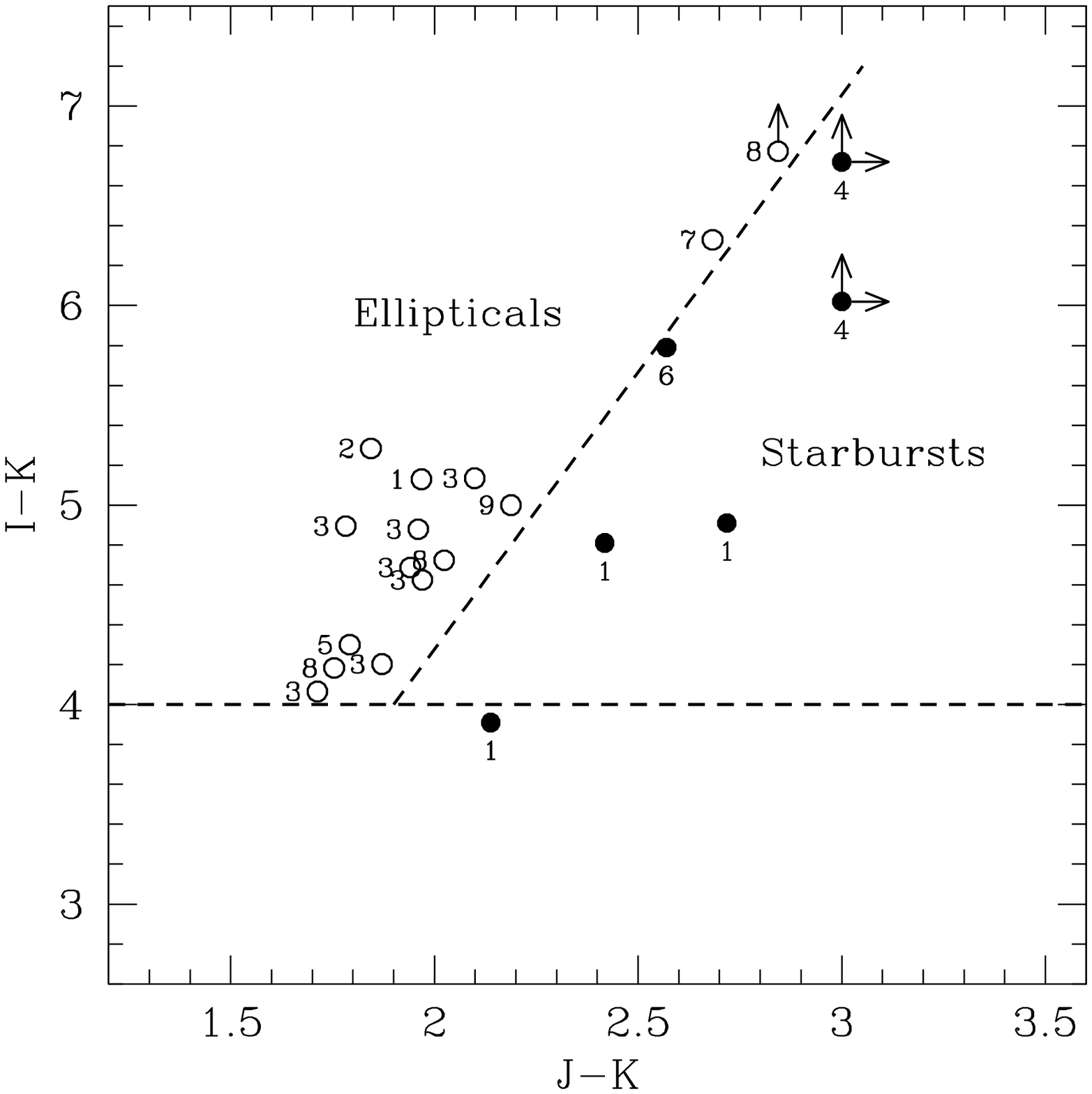,width=9cm}}
\caption{
Comparison of the selection criterion derived from the models with 
observations.
Empty dots are galaxies classified as elliptical, solid dots 
dusty starbursts. When necessary, data were transformed into the filter
system in use (Cousins' I and Bessel \& Brett J and K).
Data are labelled according to the data source:
1: Our ERGs selected in HST archive data (Cimatti et al. 1999);
2: The elliptical in the HDF-South  by Stiavelli et al. (1999);
3: Ellipticals in a X-ray selected cluster (Rosati et al. 1999);
4: Sub-mm sources detected by Smail et al. (1999) (the J magnitude of
these sources was derived from the published H mag. by assuming a
flat spectrum between these two filters, a very conservative assumption);
5 (53W091): Spinrad et al. (1997) and Bruzual \& Magris (1997);
6 (HR10): Graham \& Dey (1996);
7 (HR14): Hu \& Ridgway (1994);
8: Ellipticals selected by Benitez et al. (1999) in the HDFS with
   the photometry by Yahata et al. (2000) 
   available at http://www.ess.sunysb.edu/astro/hdfs/home.html;
9: Soifer et al. (1999).
}
\end{figure}

\section {ERGs of known nature}

These predictions can be compared with the observed properties
of ERGs with available I, J and K
magnitudes and whose nature can be independently determined
from  spectra,  morphology
or the presence of a detectable sub-mm flux.
In total we found 24 objects, 6 starbursts and 18 ellipticals,
that can be placed on the I-K vs J-K diagram .
The results are shown in Fig. 4 
(see the caption for references).
Although the number of objects 
is still small we found that all the objects fall in the 
part of the diagram predicted by the models.
Some objects of both classes fall near the separation line:
the classification in this range should be used 
with caution especially in a sample
with poor photometry.

\section {The selection criteria}
The dashed line in Figures 2, 3 and 4 which separate the two populations
for 1$<$z$<$2 in the I-K vs. J-K plane corresponds to:
\begin{equation}
(J-K) = 0.36\cdot(I-K) + 0.46~~~~~{\rm and~~~~} I-K>4
\end{equation}
and is consistent with the redshifted spectra of local objects (Fig. 2), 
the predictions of the spectral synthesis models (Fig. 3)
and the ERGs of known nature (Fig. 4).

The filter R (Cousins) could also be used instead of I to separate the 
different ERGs populations at similar redshift.
By arguments very similar to those used for Fig. 3 we predict a
separation line given by:
\begin{equation}
(J-K) = 0.34\cdot(R-K) + 0.19~~~~~{\rm and~~~~} R-K>5.3
\end{equation}
Not enough observational data are present in this colour-colour plane to
test the prediction, in particular for dusty starburst galaxies;
however cluster's galaxies at z=1.26 (Rosati et al. 1999) and
at z=1.27 (Stanford et al. 1997) with no emission line fall in 
elliptical part of R-K vs. J-K diagram, while 2 objects belonging to 
the same clusters but with [OII] emission fall in the starburst region.
Indeed at this redshift R filter is sampling the rest-frame UV region of
the spectra and is therefore more sensible than I to residual star 
formation that could affect the colour. 
I-K colour is less affected by this problem, only at higher $z$ ($>$1.8) 
ellipticals with small residual star formation could fall in the starburst
region.

We stress the point that, since $z>2$ ellipticals could fall
in the dusty region (Fig. 1), 
these colour selection criteria are valid only in $K<20-21$
selected sample, where the density of $z>2$ ellipticals 
expected in the Pure Luminosity Evolution (PLE) models 
is extremely small (Pozzetti et al. 1996).
Higher redshift ranges can be selected by using different filters;
in particular in the J-K vs. H-K diagram we derive the criterion:
\begin{equation}
(J-K) = 0.33\cdot(H-K) + 0.20~~~~~{\rm and~~~~} J-K>2.5
\end{equation}
to select old ($z_f>3$) passively evolving elliptical or
dusty starbursts at $2.0<z<2.5$.

In conclusion, the classification of the ERGs using the colour-colour
diagram I-K vs. J-K seems reliable because both model predictions and 
observed objects give the same results. 
By applying it to large complete samples it will be possible 
to remove most of the foreground stars from the ERGs samples and
statistically separate ellipticals
and starbursts, deriving important information 
to obtain their formation and evolution histories.

\section*{Acknowledgments}

We are grateful to G. Bruzual and to B. Poggianti to providing their 
spectral synthesis models.
The authors have benefit from discussion with E. Daddi, A. Cimatti,
G. Zamorani, T. Oliva, F. Basile and C. Vignali.
L.P. was supported by {\it Consorzio Nazionale per l'Astronomia e
l'Astrofisica} during the realization of this project.

\bsp

\label{lastpage}
 
\end{document}